\documentclass[aps,letterpaper,preprint,nofootinbib]{revtex4}

\usepackage{natbib}
\usepackage{cancel}
\usepackage{ulem}
\usepackage{amsthm}         
\usepackage{amsmath} 	   
\usepackage{amssymb}       
\usepackage{amsfonts}
\usepackage{graphicx} 	    
\usepackage{verbatim}
\usepackage{color}
\usepackage{float}

\definecolor{dark-red}{rgb}{0.0,0.0,0.0}
\definecolor{dark-green}{rgb}{0.0,0.0,0.0}
\definecolor{dark-blue}{rgb}{0.0,0.0,0.0}

\usepackage{enumitem}
\usepackage{multirow}

\begin{document}

\title{Scale-free primordial cosmology} 

\author{Anna Ijjas}
\email{aijjas@princeton.edu}
\affiliation{Max-Planck-Institute for Gravitational Physics (Albert-
Einstein-Institute), 14476 Potsdam, Germany} \affiliation{Rutgers 
University, New Brunswick, NJ 08901, USA}

\author{Paul J. Steinhardt}
\email{steinh@princeton.edu}
\affiliation{Department of Physics and Princeton Center for Theoretical 
Science, Princeton University, Princeton, NJ 08544, USA}

\author{Abraham Loeb}
\email{aloeb@cfa.harvard.edu}
\affiliation{Harvard-Smithsonian Center for Astrophysics, Cambridge, MA 
02138, USA}

\date{\today}

\begin{abstract}

The large-scale structure of the universe suggests that the physics 
underlying its early evolution is scale-free. This was the historic 
motivation 
for the Harrison-Zel'dovich-Peebles spectrum and for inflation.  Based on 
a hydrodynamical approach, we identify scale-free forms 
for the background equation-of-state for both inflationary and cyclic 
scenarios
and use these forms to derive predictions for the spectral tilt and 
tensor-to-scalar ratio of primordial density perturbations.  For the case 
of 
inflation, we find three 
classes of scale-free models with distinct predictions. Including all classes, we show that scale-free inflation predicts tensor-to-scalar  ratio $r > 10^{-4}$.     
We show that the observationally favored class is theoretically 
disfavored because
it suffers from an 
initial conditions problem and the hydrodynamical form
of an unlikeliness problem similar to that identified recently for 
certain inflaton potentials.   
We contrast these results with those for scale-free cyclic models.
\end{abstract}

\maketitle

\section{Introduction}

The recent \textit{Planck} satellite measurements 
\cite{Ade:2013lta,Ade:2013rta,Ade:2013ydc}, together with earlier 
observations from WMAP, ACT, SPT, and other experiments 
\cite{Sievers:2013wk}, showed with high precision that the spectrum of 
primordial density fluctuations is nearly scale-invariant, Gaussian, and 
adiabatic. These results suggest that the universe is simple and the 
physics governing its early evolution on large scales is 
`scale-free.' That is, the physics during that smoothing period in which 
the large-scale structure of the universe is determined is governed by 
dynamical equations that 
entail no dimensionful macroscopic scales and yield 
power-law solutions.  

Scale-freeness was first conjectured as a guiding cosmological principle 
over four decades ago and was the historic 
motivation for both the Harrison-Zel'dovich-Peebles spectrum 
\cite{Harrison:1969fb,Sunyaev:1970eu,Peebles:1970ag} and 
inflation 
\cite{Guth:1980zm,Linde:1981mu,Albrecht:1982wi}. In the intervening 
years, the principle seemed to lose favor as many baroque versions of 
inflationary (and other) models were proposed that explicitly introduce 
distinctive, scale-sensitive features on large scales.  The problem is 
that, without a guiding principle such as 
scale-freeness, literally any result for the spectral tilt, 
tensor-to-scalar ratio or other cosmological 
observables is possible. 
Some have emphasized this as an `attractive' feature of inflation on the 
grounds that the theory cannot be disproven  (see for 
example \cite{Ferrara:2013rsa}); but the other side of the coin is that 
this means the theory is entirely unpredictive.    

Now that scale-freeness has substantial observational support, it is 
timely to examine how this guiding principle 
dramatically collapses the range of outcomes and makes cosmological 
theories like inflation meaningfully predictive. 
We use a hydrodynamical approach that is model-independent, {\it i.e.}, 
with no reference to scalar fields or potentials, to consider the two 
well-known cosmological scenarios, the inflationary and cyclic (or 
ekpyrotic) theories of the universe. 
We identify forms for the background equation-of-state during the 
cosmological smoothing phase in each case 
consistent with {\it strict} scale-freeness. 
We also consider variations that ``weakly'' break scale-freeness. We then 
derive 
generic predictions for the spectral tilt and tensor-to-scalar ratio of 
primordial density 
perturbations resulting from the scale-free principle.

A hydrodynamical approach has been applied earlier to inflationary and 
cyclic theories \cite{Khoury:2003vb,Mukhanov:2013tua}, without explicitly 
assuming scale-freeness. 
The hydrodynamical approach is attractive since it is powerful and simple 
at the same time; it enables us to derive generic results (given the 
assumptions) and leads us to an intuitive understanding of the underlying 
physical phenomena. It is also closer to observation, in the sense that 
it is easier to determine the equation-of-state from astrophysical data 
than to determine the microphysics (scalar fields and potentials) that 
caused it. 

The goal of this paper is to show how the combination of the 
hydrodynamical approach and the principle of scale-freeness impose 
restrictions on cosmological scenarios and their 
predictions. For inflation, the combination reveals the existence of 
three distinct classes of scale-free scenarios.  We show 
that the class favored by current experiment suffers 
from an initial conditions problem and a series of other problems, 
including
 a hydrodynamic equivalent of the 
unlikeliness problem identified recently 
for certain inflaton potentials \cite{Ijjas:2013vea}. For the cyclic 
scenarios, where smoothing occurs during a period of ultra-slow 
(ekpyrotic) contraction, we find that there is only one class of 
scenarios and that 
none of the problems arise. In this analysis, we only consider a single 
contraction period without regard to whether the evolution repeats 
cyclically, so the same conclusions apply to bouncing cosmologies using 
ekpyrotic smoothing that have a single bounce or other variations.  

For the cyclic (or other ekpyrotic) theories, 
most current versions use the 
entropic mechanism 
to generate curvature perturbations \cite{Lehners:2007ac}, which
imposes the conceptual restriction that there be a two-component fluid 
to generate the perturbations. 
We find 
that handling two components rather than one in our approach is not a 
problem.  We show that scale-freeness constrains the equations-of-state 
of both components, enabling us to derive generic predictions for the 
spectral tilt and tensor-to-scalar ratio analogous to the case of 
inflation.

We believe the approach adopted here based on scale-freeness and 
hydrodynamics provides what is arguably the predictions of the simplest, best-motivated, 
and observationally best-supported models of each given cosmological theory 
and sets a standard that can be applied to any scenario in which a smooth, \textit{i.e.} scale-free background and nearly scale-invariant, adiabatic, and Gaussian perturbations are created 
at the same cosmological stage.

The paper is organized as follows. We begin in \S~II by briefly 
reviewing the inflationary and cyclic (or ekpyrotic) 
scenarios and how they can create a scale-free background.
To describe the background dynamics, in \S~III we identify forms of the 
equation-of-state consistent with the principle of 
scale-freeness for the inflationary scenario.  We demonstrate the 
existence of three distinct classes of scale-free solutions.  Then, we 
use our background solutions to derive predictions for the spectral tilt 
and tensor-to-scalar ratio of primordial density perturbations. We also 
consider cases with deviations from scale-freeness on unobservably small 
scales. 
Here and throughout the paper, our main aim is to make most generic 
statements from a minimal set of assumptions.
In \S~IV, we repeat the same type of analysis for the cyclic 
(ekpyrotic) model. 
 We conclude in \S~V by summarizing the constraints imposed by 
 scale-freeness for both the inflationary and cyclic 
theories and comparing with constraints imposed by recent data.

\section{Scale-freeness}

Both inflation and the cyclic (or ekpyrotic) theory were introduced to 
explain how inhomogeneous and anisotropic initial 
conditions  can be made  smooth and (spatially) flat, resulting in a 
scale-free universe. Inflation 
\cite{Guth:1980zm,Linde:1981mu,Albrecht:1982wi} accomplishes the feat 
with a phase of accelerated expansion occurring 
very shortly after the big bang.
Alternatively, flatness and homogeneity can be achieved by an ekpyrotic 
smoothing phase \cite{Khoury:2001wf,Khoury:2001bz}, a 
period of ultraslow contraction before the big bang.

In both phases, the dynamics can be easily understood, using a 
hydrodynamical approach in which the background 
evolution is governed by a `smoothing' fluid component (S)  with 
equation-of-state parameter, 
\begin{equation}\label{es}
\epsilon \equiv \frac{3}{2}\left(1 + w  \right)\quad \text{with}\quad
w \equiv \frac{\rho_S}{p_S}\,,
\end{equation} 
where $w$ is the equation-of-state, $\rho_S$ the energy density, and 
$p_S$ the pressure of the smoothing component. Here and throughout the 
paper we will restrict ourselves to the case that the speed of light is 
$c_s=1$.  (Although it is straightforward to extend the analysis to 
$c_s\ne 1$, current observations require $c_s > 1/3$ \cite{Ade:2013ydc}; 
for this range of $c_s$, the difference from the $c_s=1$ case is 
nominal.)
 To 
have accelerated expansion during the inflationary smoothing phase, the 
equation-of-state parameter must lie in the range $0 < 
\epsilon<1$ since the scale factor increases with time as $a \propto 
t^{1/\epsilon}$.  To have ultra-slow contraction in the ekpyrotic 
smoothing 
phase, the analogous condition is $\epsilon >3$. In both cases, 
the condition on the equation-of-state guarantees that, in the Friedmann 
equation, 
\begin{equation}\label{friedmann}
H^2 = \frac{1}{3\,M_{\text{Pl}}^2} \left( - \frac{3k}{a^2} + 
\frac{\sigma_0^2}{a^6} + \frac{\rho_S}{a^{2\epsilon}} + [\text{matter, 
radiation, etc.}] \right),
\end{equation}
the energy density in the smoothing 
component ($\rho_S \propto a^{-2\epsilon}$) can overtake  all other forms 
of energy density, including matter ($\rho \propto a^{-3}$), radiation 
($\rho \propto a^{-4}$),
and gradient energy ($\rho \propto a^{-2}$), and can also overtake the 
anisotropy ($\sigma_0^2/a^{6}$)
and spatial curvature ($k/a^{2}$).  Generally, $\epsilon\equiv 
\epsilon(N)$ is a 
function of $N$, the number of e-folds before the end of 
the smoothing phase. (Here $M_{\text{Pl}}^2 = (8\pi \mathrm{G})^{-1}$ is 
the reduced Planck mass and $\mathrm{G}$ is Newton's constant.)

In flattening the background with a single fluid of $\epsilon < 1$, 
inflation also generates a nearly scale-invariant, 
adiabatic, and Gaussian spectrum for  the curvature perturbations on 
comoving hypersurfaces characterized by a spectral 
tilt $n_s(N)-1$  \cite{Bardeen:1983qw,Mukhanov:1990me}, which is also a 
function of $N$. The same is not true for 
ekpyrosis. If there is only a single fluid in the contracting phase, the 
growing-mode, adiabatic perturbations decay and 
cannot be the seed of structure in the post-bang universe 
\cite{Creminelli:2004jg}. Currently, the best understood way of 
creating primordial density perturbations is the {\it entropic} mechanism 
\cite{Buchbinder:2007tw,Lehners:2007ac}. Here, 
pre-bang isocurvature fluctuations are generated by adding a second fluid 
component; in the simplest case, one that does not 
affect the background evolution. These isocurvature modes are 
then converted into density perturbations 
which source structure in the post-bang universe.
Another consequence of inflation is the generation of nearly 
scale-invariant tensor (gravitational wave) fluctuations.  The 
ratio of the tensor-to-scalar amplitude as a function of $N$ is labeled 
$r(N)$.  For the ekpyrotic case, the tensor amplitudes 
are exponentially suppressed compared to inflation and can be considered 
negligible for the purposes of this discussion.  
Hence, the detection or non-detection of primordial gravitational waves 
is a key means of distinguishing the two scenarios.

Assuming only that there was a period of inflation, the point has been 
made by numerous authors ({\it e.g.}, see 
\cite{Ferrara:2013rsa} for a recent example) that any observational 
outcome is possible, rendering the theory  unpredictive. 
The purpose of this paper is to use a hydrodynamical approach to 
determine how the predictions of inflationary and cyclic 
cosmologies are affected by the additional assumption that the underlying 
physics is scale-free.  
By a scale-free function we mean a
power-law form up to a coordinate-shift, {\it i.e.}, $f: 
\mathbb{R} \rightarrow \mathbb{R}$ is a scale-free function 
iff there is a coordinate transformation $\pi: \mathbb{R} \rightarrow 
\mathbb{R},\, x \mapsto x + C, \,C \in \mathbb{R}$, 
such that 
\begin{equation} \label{sf}
(f \circ \pi) (x) = \beta x^{\alpha}, \quad \alpha, \beta  \in 
\mathbb{R}.
\end{equation} 
Scale-invariant is the special case where $\alpha=0$.

For our cosmological application, we describe a cosmology as {\it 
strictly scale-free} if both the background 
equation-of-state $\epsilon(N)$ and the perturbations, characterized by 
$n_S(N)-1$ and 
$r(N)$, are scale-free.  We shall show that this 
condition is highly constraining, leading to specific predictions for 
$n_S-1$ and $r$.  
In particular, it is immediately apparent from the Friedmann equation, 
Eq.~(\ref{friedmann}), which can be written as a sum of  $a^{-
2\epsilon_i}$, that for a scale-free background the equation-of-state 
parameter of all components relevant during the smoothing stage must be 
the same.

Since the case for scale-freeness is based 
on background evolution and observations on large scales, we also 
consider {\it background-only scale-freeness} in which 
$\epsilon$ is precisely scale-free but $n_S-1$ can have deviations from 
scale-freeness on unobservably small length scales 
($N= \cal{O}(1)$).  In addition, we consider a class of models that
{\it weakly break scale-freeness} where we analyze deviations 
in $\epsilon$, $n_S-1$, and $r$ that only affect unobservably small 
scales.

\section{Inflationary theory}

In order to construct a model with ${N}^{\ast}$ {\it e}-folds of 
inflation, the following two criteria must be satisfied: 
\begin{enumerate}
\item[I:] (sufficient inflation) $N^{\ast}$ e-folds inflation occur, {\it 
i.e.}, $ \epsilon(N)  < 1$ for $1 < N < N_{\ast}$\,, and
\item[II:] (graceful exit) inflation ends in the last e-fold, {\it i.e.} 
$\epsilon(N=0) = 1$;  plus $\epsilon(N>0) < 1$ and 
$\epsilon(N<0) \geqslant 1$.
\end{enumerate} 
where $N$ is the number of $e$-folds of inflation remaining until its end  
$t_{\text{end}}$, defined as 
\begin{equation}\label{def:N}
N= \int^{t_{\text{end}}}_t H dt
\,.
\end{equation} 
$N=0$ marks the end of inflation. Here, without loss of generality we 
will assume a single continuous stage of inflation 
with $N^{\ast}$ {\it e}-folds.  If these are the only constraints 
imposed, then $\epsilon(N)$ can take many forms and the predictions can 
vary arbitrarily.  To transform inflation into a predictive theory, an 
additional constraint is needed.  We use scale-freeness as the added 
condition.

\subsection{Scale-free inflationary theory}

Scale-freeness, combined with the two numbered
criteria, determines the 
evolution of $\epsilon$ during inflation.
From Eq.~(\ref{sf}), we have 
\begin{equation}\label{esp}
\epsilon(N) = \frac{1}{(N+1)^{\alpha}},\quad \alpha >0,
\end{equation}
where $\alpha$ needs to be strictly positive to satisfy criterion I.
That is, the equation-of-state $\epsilon(N)$
consistent with the scale-free 
principle is described 
by a simple power-law form with a \textit{single} free parameter, 
$\alpha$.  The second free parameter in Eq.~(\ref{sf}), 
$\beta$, is fixed by criterion II, the condition that $\epsilon(0) = 1$. 
Considering $\beta$ as a second free 
parameter, as assumed in Ref.~\cite{Mukhanov:2013tua}, violates criterion 
II. We will discuss the implications of this 
restriction below.

To analyze different inflationary solutions, we compute the evolution of 
the Hubble parameter in terms of $\epsilon(N)$. 
Note that we need to assume {\it both} criteria I and II for this type of 
analysis. Here we are being more precise than 
some previous hydrodynamical treatments. For example, 
Ref.~\cite{Khoury:2003vb} obtains Eq.~(\ref{esp}), but through an 
inconsistent argument that first assumes  $\epsilon = \text{constant}\ll 
1$ and, hence, violates criterion II. 
In Ref.~\cite{Mukhanov:2013tua}, $\beta$ is left as a free parameter, 
which is also inconsistent with criterion II. 

For a homogeneous, isotropic, and spatially flat universe, the second 
Friedmann equation can be written as  
$\epsilon = -{\dot{H}}/{H^2}$.  Since $dN = - d\ln a$, we can rewrite the 
relation as 
\begin{equation}\label{inte}
\epsilon = \frac{d\ln H}{d\,N}\,. 
\end{equation} 
Finally, integration of Eq.~(\ref{inte}) together with our expression for 
$\epsilon$ in Eq.~(\ref{esp}) yields a closed-form 
expression for $H^2$ (or, equivalently, the smoothing 
energy density $\rho_S$) as a 
function of $N$:
\begin{equation}\label{H^2}
H^2/H_{\text{end}}^2 = \rho_S/\rho_{S,\text{end}} = \exp\left[-2 
\int^0_{N} 
\epsilon \, d\,N  \right] 
\end{equation}
which reduces in the inflationary case to
\begin{eqnarray}  \label{H^22}
H^2/H_{\text{end}}^2 &=& 
\begin{cases} (N+1)^2\,, & \text{$\alpha = 1$,}
\\
\exp\left[\frac{2\left(1- (N+1)^{1-\alpha}\right)}{\alpha-1} \right]\,, 
&\text{$\alpha \neq 1$,}
\end{cases}
\end{eqnarray}
which is the relevant observable in inflationary dynamics. Note that the 
Hubble parameter at the end of inflation, $H_{\text{end}}$, is arbitrary.

In Figure~1 we have plotted $H^2/H_{\text{end}}^2$  during 
the inflationary phase as a function of $N$ for 
different values of $\alpha$. The dashed curve corresponds with the 
strictly scale-free case, $\alpha=1$.  The rest of the 
curves are background-only scale-free.  

The curves divide into three classes: (i) the ``plateau-like" class with 
$\alpha \gtrsim 1.5$ (bold curve)  in 
which $H^2$ flattens out and is virtually independent of $N$ over the 
range  $N>60$ (changing by less than 20\%); (ii) 
the ``power-law-like" class with $\alpha \lesssim 1$ in which $H^2$  is 
unbounded above; and (iii) 
an ``intermediate class'' with $1<\alpha <1.5$, that appears power-law-like during the last 60 e-folds (see Fig.~1) but which ultimately reaches 
a plateau at very large $N\gg 60$ (with $H^2$ increasing by more than 
20\% for $N >60$).  

\begin{figure}
\begin{center}
\includegraphics[scale=0.4]{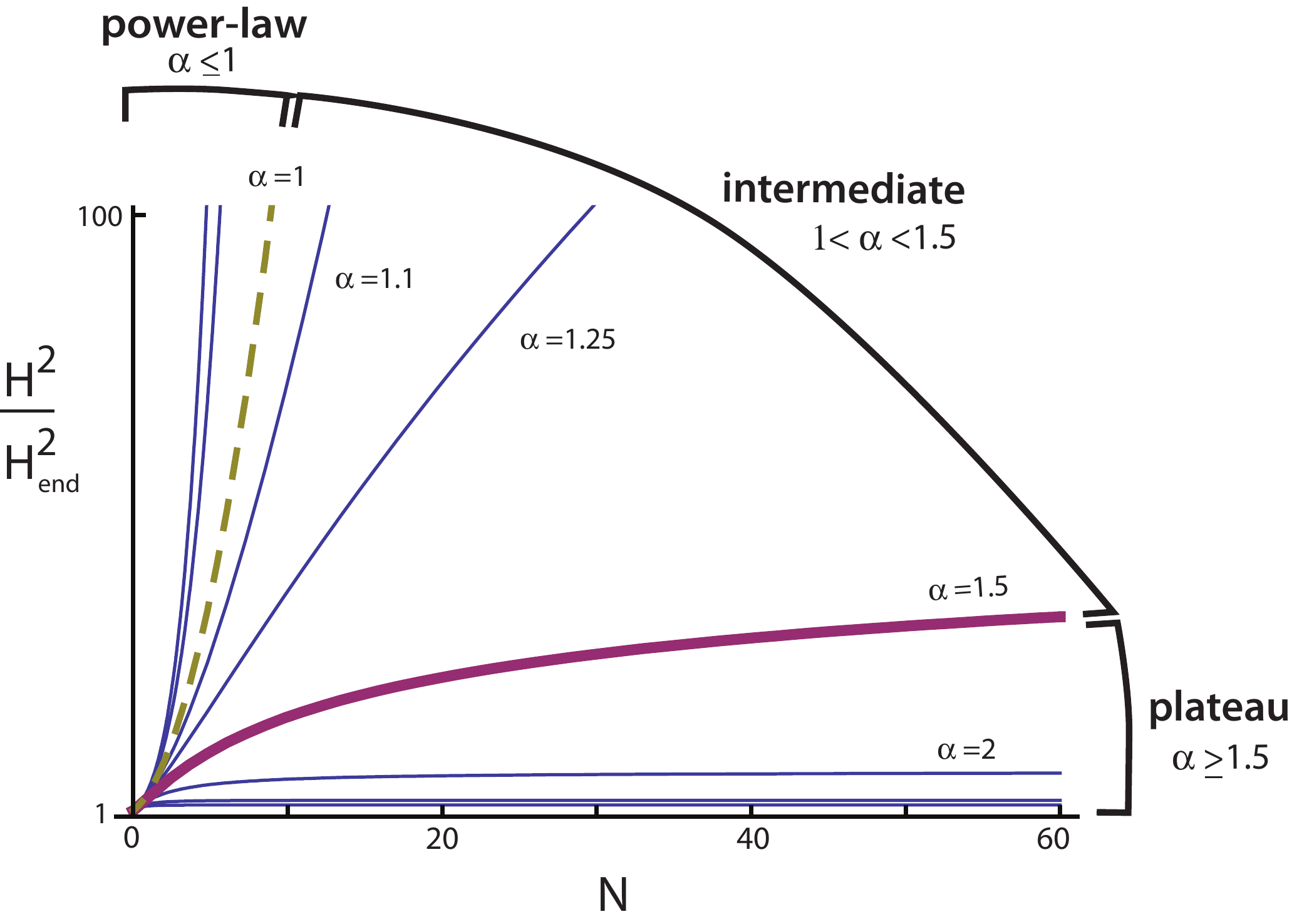}
\caption{ In the hydrodynamical picture, scale-free inflationary models 
can be divided into three classes characterized by 
$\alpha$ in Eq.~(\ref{esp}):  the {\it plateau-like} class (with $\alpha 
\ge 1.5$, where $\alpha = 1.5$ is the bold thick 
curve) in which $H^2$ flattens out 
rapidly (well before $N=60$) as $N$ increases; the {\it power law-like} 
class (with $\alpha \le 1$, where $\alpha=1$ is the dashed curve) 
in which $H^2$ is 
{\it 
unbounded} above and changes significantly as $N$ increases; and the {\it 
intermediate class} (with $1<\alpha < 1.5$), 
 which rises like a power-law for $N <60$ but which ultimately reaches a 
plateau at values of $N\gg 60$ that are irrelevant for cosmological 
predictions.  The plateau-like class is most favored by current observations but encounters the problems described in this paper.  The power law-like models are strongly disfavored by current observations but do not suffer the same problems. } 
\end{center}
\end{figure}

The expression for the equation-of-state parameter as defined in 
Eq.~(\ref{esp}) enables us to derive predictions for the 
spectral tilt and the tensor-to-scalar ratio of primordial density 
perturbations. Since $\epsilon(N)$ does not change rapidly, 
{\it i.e.},
\begin{equation}
\frac{d\ln \epsilon}{d\,N} = - \frac{\alpha}{N+1},\; \frac{d^2\ln 
\epsilon}{d\,N^2}  = \frac{\alpha}{(N+1)^2}  \lesssim 
\mathcal{O}(1)\,,
\end{equation}
we can use the approximation \cite{Wang:1997cw}:
\begin{equation}\label{n_s_i}
n_S - 1 \approx -2\epsilon + \frac{d\ln \epsilon}{d\,N}\,.
\end{equation}
Substituting $\epsilon$ from Eq.~(\ref{esp}) yields
\begin{equation}\label{n_S_inflation}
n_S - 1 \approx  - \frac{2}{(N+1)^{\alpha}} - \frac{\alpha}{N+1}\,.
\end{equation} 
It is instructive to note that $n_S - 1$ has a maximum  value of
\begin{eqnarray}\label{ns_min} 
(n_S - 1)(\alpha_0) &=& - \frac{ \ln \left[{2 (N+1) \ln (N+1)}\right] + 
1}{(N+1) \ln (N+1)}, \nonumber\\
\text{for}\quad \alpha_0 &=& \frac{\ln \left[{2 (N+1) \ln 
(N+1)}\right]}{\ln (N+1)}\,. 
\end{eqnarray} 
For example, with $N=60$, we have $\alpha_0 \simeq 1.5 $ and $(n_S - 
1)(\alpha_0) \simeq -.03$. 
This red tilt is the minimum deviation from Harrison-Zel'dovich-Peebles 
spectrum (HZP) for a scale-free inflationary model 
and is close to the observed value. (Without scale-freeness or criterion 
II, $n_S$ can be arbitrarily close to HZP or yield a 
blue-tilt.)  This extremum lies almost precisely  at the borderline 
between the intermediate and plateau-like class.  (The extremum is 
described as being at $\alpha\approx2$ in \cite{Mukhanov:2013tua}, but, 
in our analysis, this crude approximation would give the wrong impression 
that it corresponds to the observationally favored 
models deep in the plateau range when it actually corresponds 
to a disfavored case.)  

Finally, with the standard normalization, the tensor-to-scalar ratio is 
\begin{equation}\label{r}
r \approx 16\epsilon 
= \frac{16}{(N+1)^{\alpha}}\,.
\end{equation}

\subsection{Cosmological problems}

The plateau-like hydrodynamical class, especially near $\alpha=2$,  is 
the one favored by current observations \cite{Ade:2013rta}, yet it 
suffers 
from a series of problems, some of which are analogous to those described 
in the analysis of scalar field potentials in  
\cite{Ijjas:2013vea} and some of which have not been discussed 
previously:
\begin{itemize}[leftmargin=*]
\item {\it Extra parameters:}  The plateau-like class  has 
the property that  
$H^2$ is nearly flat except for the last e-fold or so when the 
expansion rate suddenly decreases; 
see the feature at small $N$ in the plateau-like curves in Fig.~1. 
That means whatever 
microphysics accounts for $\epsilon(N)$ must have an extra
parameter and/or field compared to the power-law-like models 
adjusted to rapidly cutoff the 
inflation after a long period of a nearly constant $H^2$.  
We will see this effect in \S~V when we translate our hydrodynamical 
results into models of scalar-fields and inflaton potentials.
\item
{\it Hydrodynamical initial conditions problem:}  As originally imagined, 
inflation was supposed to smooth and flatten the universe beginning from 
arbitrary initial conditions after the big bang 
\cite{Guth:1980zm}.  However, this view had to be abandoned as it was 
realized that large inflaton kinetic energy and 
gradients prevent inflation from starting.  Consequently, inflation can 
only take hold if the entropy, kinetic energy, and 
gradients within a Hubble-sized patch is exceedingly small.  

We note that the 
later that inflation starts, the greater is the physical size of 
a Hubble patch and the more unlikely is the initial condition. 
A distinctive feature of the power law-like hydrodynamic class ($\alpha 
\le 1$)  is that $H^2$ is unbounded 
above.  Hence, inflation can begin, in principle, at arbitrarily high 
$H^2$ or, equivalently, over a small patch where the initial 
conditions are less unlikely compared to cases where inflation starts 
later.   This includes inflation beginning immediately 
after the big bang when the energy density is at the Planck scale.  

By contrast, inflation for models in which $H^2$ is bounded above, ({\it 
i.e.}, all $\alpha>1$),  can only begin after the universe expands enough 
for the energy density to drop to the 
level of the plateau, $M_{\text{I}}^4$. The Planck2013 constraint on $r$ 
($r_{0.002} < 0.12$ at 95\% CL) \cite{Ade:2013rta} yields
\begin{equation}
M_{\text{I}}^4 \lesssim  \frac{3\pi^2A_s}{2}\,r\,M_{\text{Pl}}^4\sim 
10^{-12}\,M_{\text{Pl}}^4\,\frac{r_{\ast}}{0.12}
\end{equation}
at 95\% CL, where $A_s$ is the scalar amplitude and $r_{\ast}$ the value 
of  $r$ evaluated at Hubble exit during inflation of mode with wave 
number $k_{\ast}$.  
This is well below the Planck density at a time 
when the Hubble volume is, by simple comparison of the scales 
$M_{\text{Pl}}/M_{\text{I}} \sim 
10^3\cdot(10^{16}\,\text{GeV}/M_{\text{I}})$, a billion times (or more) 
greater  \cite{Ijjas:2013vea}.
In this case, some combination of gradient energy 
density, spatial curvature, and radiation must necessarily dominate 
immediately after the big bang and for a substantial period thereafter 
before inflation can ever take hold.  A well-known problem, 
though, is that gradient energy and spatial curvature tend to block 
inflation by causing regions of space to collapse before 
inflation can start \cite{Ijjas:2013vea}.  That is, inflation can only 
begin for the plateau-like models if there is the extraordinary 
additional assumption that the universe emerges from the big bang with a 
patch,
\begin{eqnarray}
R^3(t_{\text{Pl}}) &\gtrsim& \left[ a(t_{\text{Pl}}) 
\int^{t_{\text{I}}}_{t_{\text{Pl}}} \frac{d\,t}{a}\right]^{3} \sim 
\left[\frac{a(t_{\text{Pl}})\,H(t_{\text{Pl}})}{a(t_{\text{I}})\,H(t_{\text{I}})}H^{-1}(t_{\text{Pl}})\right]^{3}\nonumber \\ 
&>& 10^{9} \left( \frac{10^{16}\,\text{GeV}}{M_{\text{I}}} 
\right)^3  
H^{-3}(t_{\text{Pl}}),
\end{eqnarray}
that is smooth 
and flat on scales a billion times greater than 
required for the unbounded power-law-like case \cite{Liddle:2000cg}.   
Our hydrodynamic 
analysis divides the inflationary models along the 
dashed line ($\alpha =1$) in Fig.~1 between those that require this 
extraordinary assumption (plateau-like and intermediate with $\alpha >1$) 
and those that do not ($\alpha \le 1$).
\item
{\it Hydrodynamical unlikeliness problem:}   Even assuming the rare 
initial 
conditions are satisfied, the observationally favored 
plateau-like models ($\alpha \approx 2$) produce exponentially less 
smooth and flat volume than the power-law-like or intermediate class 
models with $1 \lesssim \alpha <1.5$. 
This leads to the hydrodynamic version of the ``unlikeliness problem'' 
similar to (but not identical to; see \S~V) the one discussed in 
\cite{Ijjas:2013vea}:  In most energy 
landscapes with plateau-like inflation, even relatively simple 
landscapes, 
there are also paths for inflation to proceed to the same vacuum
 with power-law-like 
inflation.  A simple example discussed in \cite{Ijjas:2013vea} is ``new 
inflation,'' where power-law is favored by a factor of $\exp(10^8)$ 
\cite{Ijjas:2013vea}.  The fact that the same current state can be 
reached 
by either route but plateau-like produces exponentially less volume makes 
the plateau path exponentially less likely based on theoretical 
reasoning.  The fact that observations currently favor a plateau path is 
problematic.

In order to determine the maximum inflated volume possible for each 
hydrodynamic model, we must determine the largest 
value of $N$ for which  the density fluctuation $\delta \rho/\rho (N)$ is 
less than 1.   For larger $N$ where $\delta 
\rho/\rho$ exceeds 1,  quantum fluctuations totally spoil the homogeneity 
and curvature.  Hence, $N_{\text{max}}(\alpha)$, the  
maximum number of e-folds as a function of $\alpha$, is determined by the 
condition 
\begin{equation}\label{def}
\delta\rho/\rho\,(N_{\text{max}}) = 1 .
\end{equation}
The fluctuation amplitude is 
\begin{equation}\label{exp}
\delta\rho/\rho\, (N) \simeq \frac{H(N)}{M_{\text{Pl}}\sqrt{\epsilon(N)}}
 \end{equation}
(for the derivation use, for example, $\delta\rho/\rho = H / \dot{\phi}$ 
and $\dot{\phi}^2 = \rho + p$).
Substituting the expressions we previously found for $H^2$ and 
$\epsilon$, Eq.~(\ref{def}) and (\ref{exp}) together give
\begin{equation}
N_{\text{max}}(\alpha) = 
-1 + \left( \frac{1}{2}\,\alpha\, W(z) \right)^{\frac{1}{1-\alpha}}\,,
\end{equation}
where $W$ is the Lambert $W$ function, and its parameter
\begin{equation}
z = \frac{2}{\alpha}\left( 10^5\cdot 61^{\alpha/2}\cdot 
\exp\left(\frac{61^{1-\alpha}}{1-\alpha}\right) 
\right)^{\frac{2}{\alpha}(1-\alpha)}\,,
\end{equation}
and $\delta\rho/\rho$ is normalized such that $\delta\rho/\rho(N=60) = 
10^{-5}$. 
 For $\alpha=1$, $N_{\text{max}}(\alpha)$ is $61\cdot 10^{10/3}\approx 
10^5$.   
\begin{figure}
\begin{center}
\includegraphics[scale=0.45]{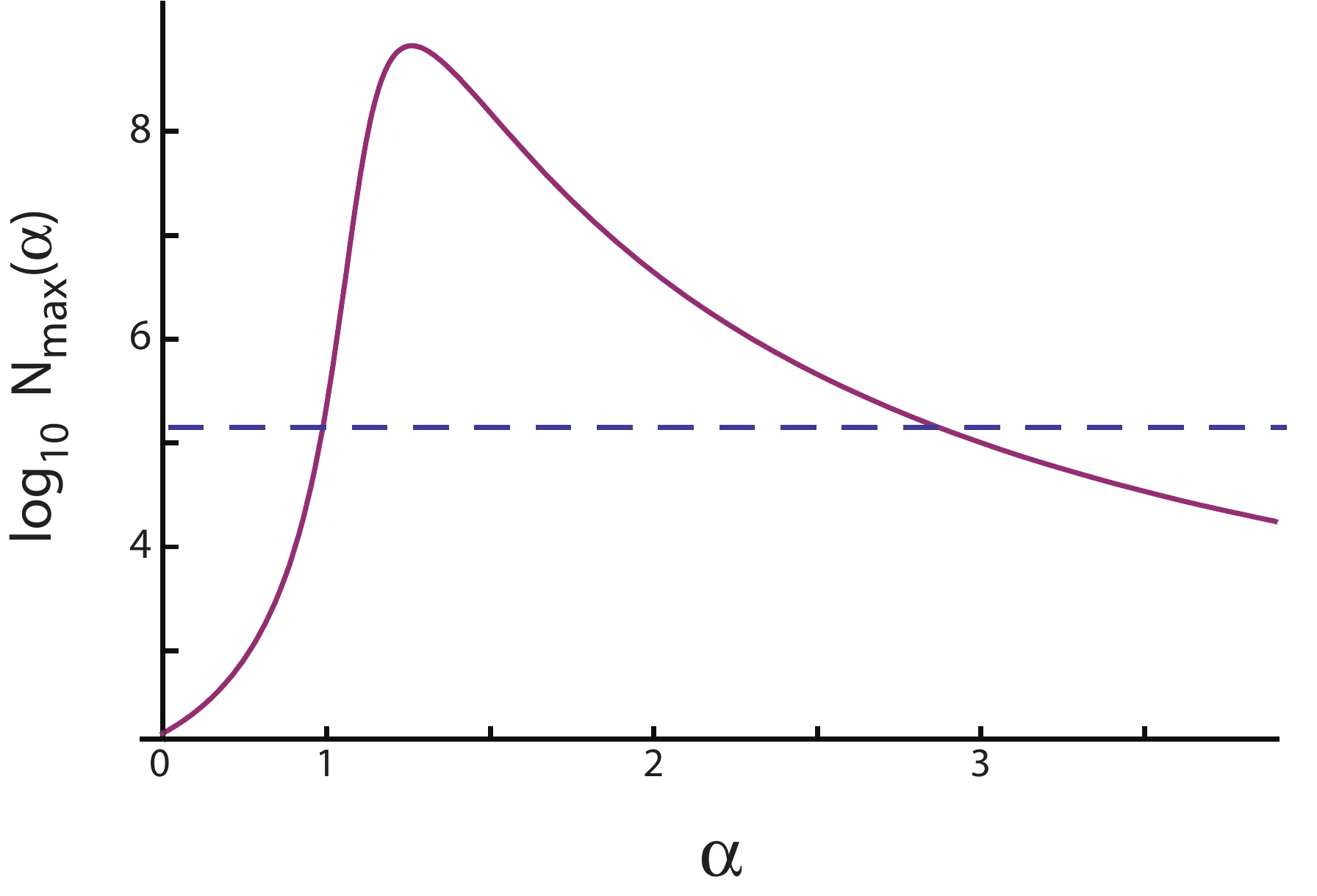}
\caption{ A logarithmic plot of the maximum number of e-folds 
$N_{\text{max}}(\alpha)$ for scale-free models as a function of the 
hydrodynamic variable $\alpha$. The plot assumes initial conditions can 
be set perfectly smoothly in the initial Hubble 
patch.  } 
\end{center}
\end{figure}

As illustrated in Fig.~2, $N_{\text{max}}$ is maximal overall for 
$\alpha\simeq 1.25$; among the power-law-like cases, $\alpha=1$ is most 
favored; and among the plateau-like models $\alpha=1.5$ is most favored. 
The differences in 
inflated volume in each case are exponentially large, of order ${\rm exp} 
(10^{5-8})$, so ``favored'' means ``very strongly 
favored'' \cite{Ade:2013rta}.  Note that $\alpha=2$ is strongly 
disfavored; yet, 
this is the 
inflationary type model that is currently most favored observationally.  
 
These estimates for $N_{\text{max}}(\alpha)$ are, however, overly 
optimistic, assuming that the initial 
conditions when the universe emerged from the big bang
could be set with arbitrary accuracy so that the energy density in the 
smoothing component is the maximum possible, $3 
H^2(N_{\text{max}}(\alpha))$ 
in Planck units. In fact, as we pointed out above, most patches of space 
are likely to have large gradient energy that will spoil inflation 
altogether.  Even if we eliminate those patches and consider only 
homogeneous patches, in each patch there remain different mixes of 
radiation, kinetic energy, potential energy, and other forms of energy 
such that, typically, we do not have patches at precisely the ideal 
potential energy to obtain $N_{\text{max}}$. Hence we should imagine some 
flex of order $x$ in the amount of the initial potential energy. 
A reduction of the average 
energy density in the patch by a factor $x$ requires a revised estimate 
$N_{\text{max}}(\alpha,x)$: 
\begin{equation}
N_{\text{max}}(\alpha,x) = 
\left( N_{\text{max}}(\alpha,0)^{1-\alpha}  - \frac{\alpha-1}{2}\ln (x) 
\right)^{\frac{1}{1-\alpha}} - 1,   
\end{equation}
which equals $61\cdot 10^{10/3} \sqrt{x}$ for $\alpha = 1$.   Because 
plateau-like models with $\alpha \ge 1.5$ are so flat for large $N$, a 
reduction in average $H^2$ by some factor $x$ produces a much greater 
reduction in $N_{\text{max}}(\alpha,x)$ relative to 
$N_{\text{max}}(\alpha)\equiv N_{\text{max}}(\alpha,0) $ than is found 
for power-law-like models.

\begin{figure}
\begin{center}
\includegraphics[scale=0.535]{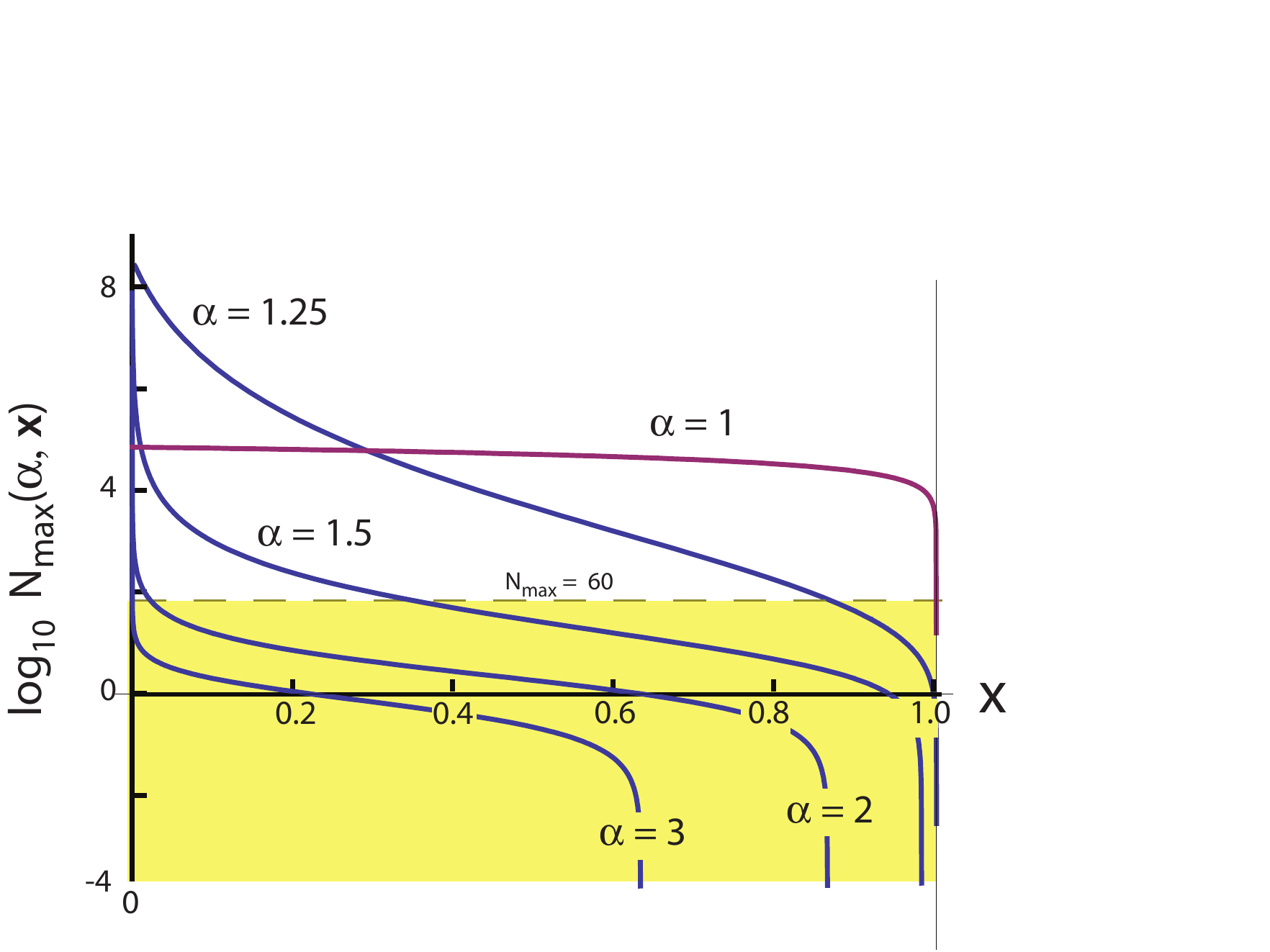}
\caption{The sensitivity of $N_{\text{max}}$ to the initial energy 
density in the smoothing component at the Planck time when the universe first emerges from the big bang.  If the energy density in a patch could be set with perfect precision, the maximum number of $e$-folds of inflation 
would be $N_{\text{max}}(\alpha)\equiv  N_{\text{max}}(\alpha,0)$ plotted 
in Fig.~2.  Due to contributions of other forms of energy (kinetic 
energy, radiation energy, {\it etc.}), we assume a variation of $x$ percent 
from perfect precision and compute how this affects the maximum number of 
$e$-folds, $N_{\text{max}}(\alpha,x)$, as shown in the logarithmic plot above.    
Note that the $N_{\text{max}}(\alpha)$ in Fig.~2 is equal to 
$N_{\text{max}}(\alpha,0)$. The plot shows that 
$N_{\text{max}}(\alpha,x)$ for $\alpha=1$ (strictly scale-free power-law-like models) is rather insensitive to $x$. By contrast, plateau-like models ($\alpha \ge 1.5$) are so extremely sensitive to $x$ that, unless the initial energy density of the smoothing component is  set with extraordinary precision, the value of $N_{\text{max}}(\alpha,x)$ is much less than that for the power-law-like class  and less than the minimal 60 needed for inflation. The shade region corresponds to insufficient inflation.} 
\end{center} \end{figure}

Fig.~3 shows  $\log N_{\text{max}}$ as a 
function of $x$ for different values of $\alpha$. The dashed 
line corresponds to the strictly scale-free, unbounded power-law-like case with $\alpha=1$; the thin black curves to models with $\alpha$-values of $1.25,1.5,2, 3$; 
the red horizontal line 
marks 60 $e$-folds. It is clear that the plateau-like 
models fail to reach $N=60$ $e$-folds for even a small  $x$, while the power law-like models and intermediate class models are comparatively insensitive to the initial distribution of energy in the patch.  

In sum, there are three classes of scale-free inflationary scenarios. Power-law-like models
 ($\alpha \le 1$) do not suffer from  the initial conditions problem or unlikeliness problem.  Models of the 
intermediate class have the initial conditions problem, but not the 
unlikeliness problem.  However, these models are all observationally 
disfavored currently \cite{Ade:2013rta}.
The observationally-favored plateau-like models with $\alpha=2$ 
suffer from all the problems described above. 
Hence, the theoretically favored scale-free inflationay models are 
observationally disfavored and vice versa.
The fact that the initial 
conditions and unlikeliness problems impose different constraints 
illustrates that they are logically distinct, a point that some have 
disputed in 
discussions of \cite{Ijjas:2013vea}.
\end{itemize}

\subsection{Deviations from scale-freeness}

We have thus far considered $\epsilon(N)$ that have a scale-free form. 
The case $\alpha=1$ is {\it strictly} scale-free in that the functions 
that describe the background, $\epsilon(N)$ and $H(N)$, as well as the 
functions that describe the perturbations 
\begin{equation}
n_S (N) -1 =  \frac{3}{N+1}\,
\end{equation}
are all simple power-laws (or power-laws with shifts).  

For $\alpha \neq 1$, the background functions are still scale-free but 
the spectral index is not: 
\begin{equation}\label{dev}
n_S (N) - 1= - \frac{2}{(N+1)^{\alpha}}- \frac{\alpha}{N+1}
\end{equation}
so there is only {\it background} scale-freeness. 

For {\it weakly broken} scale-freeness, there can be no complete 
treatment since ``weakly" is an imprecise term.  Here we consider in this 
category deviations from scale-freeness 
at the background level but only on length scales that are 
unobservably small (corresponding to small $N$):  

\begin{equation}\label{nsf1}
\epsilon = \frac{\beta}{(N+1)^{\alpha}} + \frac{\beta-
1}{(N+1)^{\alpha+\gamma}}\,, \quad \text{with} \quad \beta, \gamma > 0,\; 
\beta \neq 1\,,
\end{equation}
where this form is designed to still satisfy inflationary criteria I and 
II. 
For the deviation to be small, in addition, it is necessary that 
\begin{equation}\label{gamma}
|1-1/\beta| \ll  (N+1)^{\gamma}\quad \text{and} \quad |\beta - 1| < 1
\end{equation}
for observable $N$. 
Then, with an additional free parameter, the 
predictions are modified:
\begin{equation}\label{nsf2}
\epsilon \approx \frac{\beta}{(N+1)^{\alpha}}\,, \quad n_S -1 \approx 
\begin{cases} 
- \frac{2\beta}{(N+1)^{\alpha}}\,, & \text{$\alpha < 1$,}\\
-\frac{2 \beta+ 1}{(N+1)}\,, & \text{$\alpha = 1$,}\\
- \frac{\alpha}{N+1}\,, &\text{$\alpha > 1$,} 
\end{cases} \quad r \approx \frac{16\,\beta}{(N+1)^{\alpha}}\,.
\end{equation}

As we shall discuss below in \S~V, the case $\alpha = 1$ is of particular 
interest as it corresponds to power-law inflaton ($\phi$) potentials 
$V(\phi) \propto \phi^n$ with  $n= 4\beta$.  From Eq.~(\ref{nsf2}), we 
note 
that the weakly scale-free breaking inflationary models ($\beta \neq 0$) 
entail 
two independent parameters while strictly scale-free inflationary theory 
involves exactly one free parameter.

\section{Cyclic theory}

In the following section, we carry out the same type of hydrodynamical 
analysis for the 
cyclic theory that we previously did for inflation. 
In order to construct a model with $\mathcal{N}^{\ast}$ $e$-folds of 
ultra-slow contraction (ekpyrosis) that flattens and smoothes the 
universe, the two criteria analogous to those used for inflation are as 
follows:
\begin{enumerate}
\item[I$'$:](sufficient ekpyrosis) $\mathcal{N}^{\ast}$ $e$-folds of 
ekpyrosis 
occur, {\it i.e.}, $\epsilon(\mathcal{N}) > 3$ for $1 < \mathcal{N} < 
\mathcal{N}^{\ast}$; and
\item[II$'$:] (exit) ekpyrosis ends in the last $e$-fold, {\it i.e.}, 
$\epsilon(\mathcal{N}>0) > 3$, and $\epsilon(0) = 3$.
\end{enumerate}
We have introduced the dimensionless time variable $\mathcal{N}$, defined 
by
\begin{equation}
\mathcal{N} \equiv \ln \left( \frac{a_{\text{end}}\, 
H_{\text{end}}}{a\,H} \right)\,.
\end{equation}
$\mathcal{N}$ measures the number of $e$-folds of modes that exit the 
horizon before the end of ekpyrosis. It is related to the time variable 
$N$ used in the previous section by $d\,\mathcal{N} = 
(\epsilon-1)\, dN$.  For inflation $\mathcal{N} \approx N$, 
since $H \approx $ constant during accelerated expansion. For ekpyrosis, 
on the other hand, $\mathcal{N} \gg N$ because $H$ grows
significantly during ultra-slow contraction while $a$ shrinks very 
slowly, so $\mathcal{N}$ is the correct time-variable to use.  
Here, in analogy with the treatment of inflation, we assume a single 
continuous stage of 
ekpyrosis with $\mathcal{N}^{\ast}$ $e$-folds.

\subsection{Scale-free cyclic theory}

Scale-freeness, combined with these two criteria, determines the 
evolution of $\epsilon$ during the ekpyrotic phase. From Eq.~(\ref{sf}) 
together with criteria I$'$ and II$'$, we have
\begin{equation}\label{eqp}
\epsilon(\mathcal{N}) = 3\,(\mathcal{N} + 1)^{\alpha_1}, \quad \alpha_1 >0.
\end{equation}
That means, the shape of the equation-of-state parameter consistent with 
the scale-free principle is a simple power-law form with a single free 
parameter. The second free parameter, $\beta_1$, in Eq.~\ref{sf} is fixed 
by criterion II$'$, which requires $\epsilon(0)=3$. 

To analyze different cyclic solutions, we study the evolution of the 
total energy density $H^2/H^2_{\text{end}}$ during ekpyrosis. 
Substituting Eq.~(\ref{eqp}) into Eq.~(\ref{H^2}) yields 
\begin{equation}\label{H_cyclic}
H^2/H^2_{\text{end}} = \exp\left( - 2\,\mathcal{N} + 
2\int^0_{\mathcal{N}}\frac{d\mathcal{N}}{3(\mathcal{N}+1)^{\alpha_1}-1} 
\right).
\end{equation}
Note that this expression reflects a characteristic feature of an 
ekpyrotic phase that $H^2$ grows by many orders of magnitude during 
smoothing.  Figure~{4} shows a logarithmic plot of $H^2/H^2_{\text{end}}$ 
for the 
ekpyrotic phase as a function of $\mathcal{N}$ for different values of 
$\alpha_1$. 

\begin{figure}
\begin{center}
\includegraphics[scale=0.525]{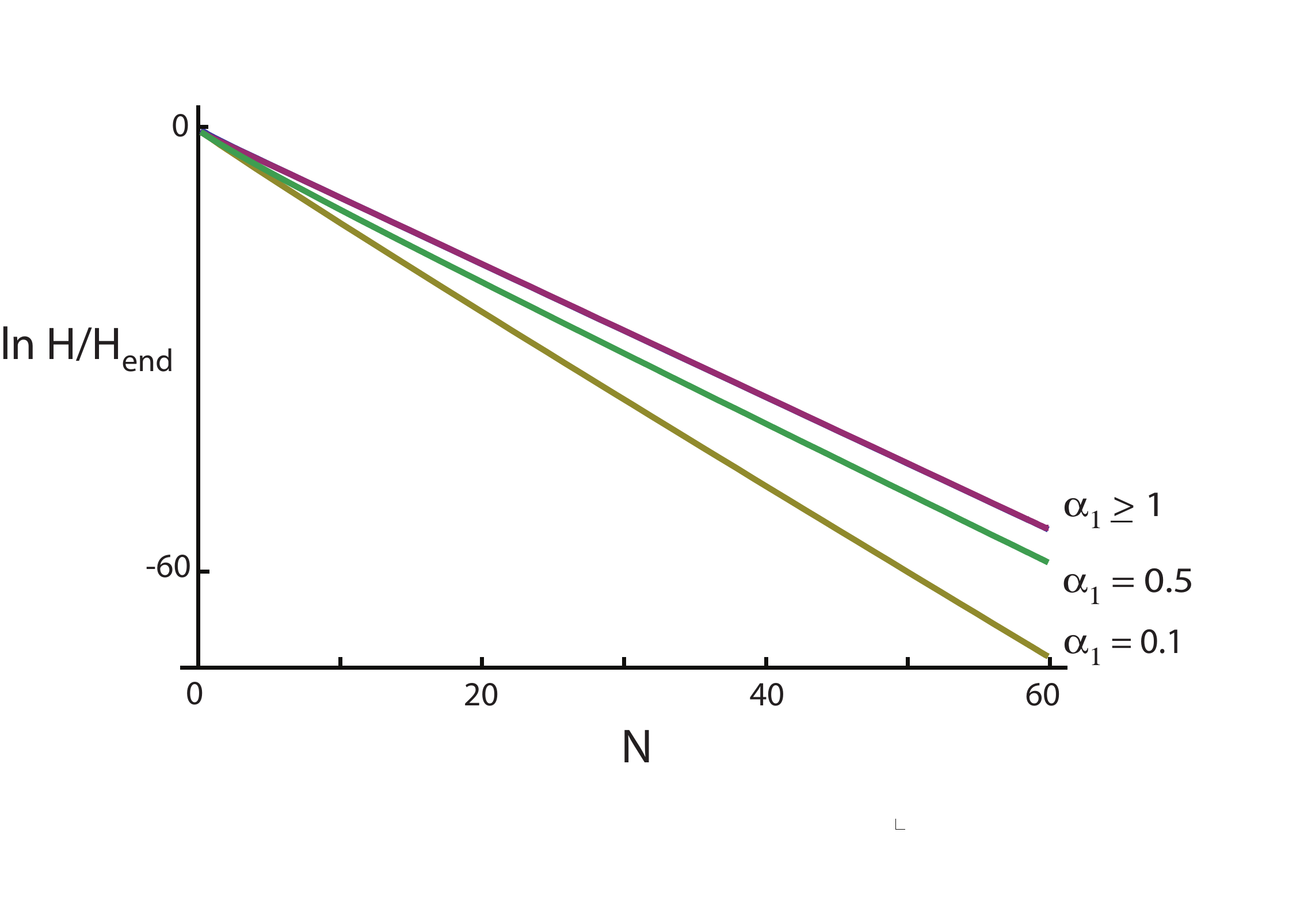}
\caption{Plot of $\ln H^2/H^2_{\text{end}}$ vs. $\mathcal{N}$ for the cyclic picture 
for a range of $\alpha_1$.}           
\end{center}\end{figure}

In contrast to inflation, cyclic models do not divide into different 
classes. In fact, for $\alpha_1 \gtrsim 1$ 
all of the $H^2$ curves lie virtually on top of one another such that the 
Hubble parameter proves 
effectively independent of $\alpha_1$. Hence, the unlikeliness problem, 
based on comparing the probability of different classes, cannot 
arise for the cyclic theory. In addition, it follows 
from the $\alpha_1$-independence 
 that choosing  a value of $\alpha_1$ to 
fit observational data does not involve any special fine-tuning relative 
to the general class of models.

The initial conditions requirement is extremely mild.  It suffices to 
have a volume of space on the scale of meters in diameter that is absent 
of black holes or non-linear structure at the beginning of the 
contraction phase \cite{Erickson:2006wc}.  The ekpyrotic mechanism will 
smooth and flatten this region and the bounce will transform this region 
during the expansion phase into a size of order the Hubble volume today.  
The initial condition can be reached in a number of ways, including by 
having an expanding phase precede the contraction phase. For example, in 
the cyclic scenario, the initial condition is easily achieved by having 
the ekpyrotic phase preceded by an expanding dark 
energy dominated phase just like the current phase of our universe. 
Consider that the present universe already contains exponentially many 
patches that satisfy the initial condition requirements and 
any further expansion only increases their number.  Having an expanding 
dark energy phase turn into a contracting phase is known to be quite 
straightforward to achieve, {\it e.g.}, by having a scalar field roll or 
tunnel from a phase with positive potential density to a phase  with a 
negative potential energy density 
\cite{Steinhardt:2002ih,Steinhardt:2001st}.
In order 
for ekpyrosis to occur, no further criteria need to be satisfied; 
expansion can turn into contraction at arbitrarily low energies for an 
$\alpha_1$ since there is no (classical) limit in Fig.~4 on how low $H$ can be when contraction begins for any $\alpha_1$ (so the choice of $\alpha_1$ 
does not require extra tuning).  By contrast, for inflation, assuming an 
expanding phase after the 
bang is not sufficient since the natural conditions after the bang would 
have large gradient and kinetic energies that would block the initiation 
of inflation.

In sum, at background level, none of the problems pointed out above for 
inflation arise for the cyclic model. There is no fine-tuning or 
unlikeliness problem, and there is no initial conditions problem 
comparable to the inflationary case.  

At the perturbative level there is a notable conceptual difference 
between inflation and the cyclic model, at least according to most 
current versions of cyclic theory.  Namely, the generation of primordial 
density  perturbations is assumed to be a two-stage process. First, 
entropy or isocurvature perturbations are created before the bounce. 
These perturbations are then converted into primordial density 
perturbations at some time during the transition from big crunch to big 
bang \cite{Lehners:2007ac}. 

Modeling this scenario in a hydrodynamical approach requires a two-component fluid: one fluid component governs the background evolution and 
the other is responsible for the generation of isocurvature fluctuations. 
The background fluid component can be described by an equation-of-state 
parameter, $\epsilon_1(\mathcal{N})$, as defined in Eq.~(\ref{eqp}), 
\begin{equation}\label{e_1}
\epsilon_1(\mathcal{N}) = 3\,\beta_1 (\mathcal{N} + 1)^{\alpha_1}, \quad 
\alpha_1>0\,,
\end{equation}
where $\beta_1=1$ according to criterion II$'$.
The equation-of-state parameter for the second fluid, 
$\epsilon_2(\mathcal{N})$, must also satisfy the requirement of scale-
freeness. Hence, from Eq.~(\ref{sf}), it is necessary (but not 
sufficient, as we point out below) for $\epsilon_2(\mathcal{N})$ to take 
the form
\begin{equation}\label{e_2}
\epsilon_2 (\mathcal{N}) = 3\,\beta_2\,(\mathcal{N} + 1)^{\alpha_2}, 
\quad \alpha_2 \in \mathbb{R}\,.
\end{equation}
If this component satisfies the null energy condition, $\beta_2$ must be 
greater than or equal to zero.
 
Before imposing any further conditions, the general expression for the 
spectral tilt of density perturbations is
\begin{equation}\label{n_s_cyclic0}
n_S(\mathcal{N}) - 1 = 3-\sqrt{1+8\kappa}\left( 1+ 3 \cdot \frac{1-
2\kappa}{1+8\kappa}\cdot\frac{2}{\epsilon_1} + \frac{8-
5\kappa}{1+8\kappa}\cdot \frac{\epsilon_1,_{\mathcal{N}}}{\epsilon_1} 
\right)
\,,
\end{equation}
where  
\begin{equation}\label{kappa}
\kappa(\mathcal{N}) = \epsilon_2 / \epsilon_1 = (\beta_2/\beta_1)\,( 
\mathcal{N} + 1 )^{\alpha_2-\alpha_1}
\end{equation}
(see the Appendix for the derivation).
In the limit of constant $\kappa(N) \approx 1$, the expression reduces to
\begin{eqnarray}\label{n_s_cyclic00}
n_S - 1 &=&  \frac{2}{\epsilon_1} - 
\frac{\epsilon_1,_{,\mathcal{N}}}{\epsilon_1} + \frac{4}{3}(1-\kappa)\,,
\end{eqnarray}
in agreement with \cite{Lehners:2013cka,Buchbinder:2007tw}.

\subsection{Deviations from scale-freeness}

For the {\it strictly} scale-free case, both the background and the 
perturbations must be simple power-laws.  For the background Friedmann 
equations, we mean that the dominant contribution to $H^2$ in 
Eq.~(\ref{friedmann}) should be a simple power-law in $a$. As noted above 
in Eq.~(\ref{sf}), this requires 
$\epsilon_1(\mathcal{N})=\epsilon_2(\mathcal{N})$ with 
$\alpha_1=\alpha_2=1$ and $\beta_1=\beta_2$. Then, the prediction for the 
spectral tilt is 
\begin{eqnarray}\label{n_s_cyclic00}
n_S - 1 &=&  \frac{2}{\epsilon_1} - 
\frac{\epsilon_1,_{,\mathcal{N}}}{\epsilon_1} \nonumber\\
&=& - \frac{1}{3\,(\mathcal{N} + 1)}
\,.
\end{eqnarray}

For the {\it background-only} scale-free case, we still require 
$\beta_2=\beta_1=1$ and $\alpha_1=\alpha_2$, but the $\alpha$'s need not 
be 1.  Then, the spectral tilt has a small deviation from scale-freeness;
\begin{eqnarray}\label{n_s_cyclic}
n_S - 1 &=&  \frac{2}{\epsilon_1} - 
\frac{\epsilon_1,_{,\mathcal{N}}}{\epsilon_1} \nonumber\\
&=& \frac{2}{3\,(\mathcal{N} + 1)^{\alpha_1}} - 
\frac{\alpha_1}{\mathcal{N} + 1}
\,,
\end{eqnarray}
in agreement with \cite{Buchbinder:2007tw,Lehners:2007ac}.  Note that, 
even though there are two fluid components, the expression for $n_s$ has 
only one free parameter, as in the case of inflation.

Finally, we consider the {\it weakly broken} scale-free case in which 
deviations from scale-freeness occur only on unobservable scales.  As 
with inflation, there is no absolute definition of weakly broken scale-free, but we consider two types of deviations that arise in microphysical 
models of scalar fields. 

First, a very weakly broken scale-free background occurs if $\beta_2$ is 
close to but not equal to $\beta_1=1$, or, equivalently,  $0<|\kappa -
1|\ll1$. 
In this case, the expression for the tilt reduces to the simpler form
\begin{eqnarray}\label{n_s_cyclic}
n_S - 1 &=& \frac{4}{3}(1-\kappa) + \frac{2}{\epsilon_1} - 
\frac{\epsilon_1,_{,\mathcal{N}}}{\epsilon_1} \nonumber\\
&=& \frac{4}{3}(1-\kappa) + \frac{2}{3\,(\mathcal{N} + 1)^{\alpha_1}} - 
\frac{\alpha_1}{\mathcal{N} + 1}
\,,
\end{eqnarray}
in agreement with \cite{Buchbinder:2007tw,Lehners:2013cka}.
A second type of deviation from background scale-freeness is to choose 
$\beta_1 \ne 1$, which generates additional contributions to $n_S$  
analogous 
to the inflationary case; see Eqs.~(\ref{nsf1}) and (\ref{nsf2}).  As 
with the background case, the weakly broken scale-free case for the two 
fluid-component cyclic scenario has the same number of free parameters as 
for inflation, so neither theory is advantageous by this measure.   

\section{Scale-free scalar fields and potentials}

The problems we identified for inflationary theory are similar to but not identical to the issues 
identified previously in \cite{Ijjas:2013vea}, using a model dependent 
analysis based on assuming that inflation is driven by scalar fields with 
specific potential forms.
In order to compare the two approaches, we translate our 
hydrodynamical scale-free models into the field picture, first for 
inflation and subsequently for cyclic cosmology.

\subsection{Scale-free inflationary potentials}

The construction of scale-free inflationary potentials corresponding to 
the hydrodynamical models described in previous sections is based on 
assuming single-field, slow-roll inflation with canonical kinetic energy 
density and $\rho_S \simeq V(\phi)$, where $V(\phi)$ is the potential 
energy density for the inflaton scalar field $\phi$. 
Following the method presented in  \cite{Mukhanov:2013tua}, the Friedmann 
equations together with the identity $\dot{\phi}^2 = \rho_S + p_S$ yield 
\begin{equation}\label{Nphi}
\frac{\phi - \phi_{\text{end}}}{M_{\text{Pl}}} = \pm \int^0_N 
\sqrt{2\epsilon}\, d\,N = \pm \sqrt{2} \cdot \begin{cases} - \ln(N+1), & 
\alpha = 2\\
\frac{2}{2 - \alpha}\left( 1 - (N+1)^{\frac{2 - \alpha}{2}} \right), & 
\text{otherwise}. \end{cases}
\end{equation}
Then, with Eq.~(\ref{H^2}) we find the expression for the inflationary 
potential
\begin{equation}\label{Vphi}
V(\phi)  =  \begin{cases} \lambda\,\phi^4, & \alpha = 1,\\
V_{\text{end}}\,\exp \left( 2 - 2\,\exp\left(- \frac{\phi - 
\phi_{\text{end}}}{\sqrt{2}\, M_{\text{Pl}}} \right) \right), & \alpha = 
2, \\
V_{\text{end}}\,\exp \left( \frac{2}{1 - \alpha}\left( \left( 1 \pm 
\frac{2 - \alpha}{2\sqrt{2}}\cdot \frac{\phi - 
\phi_{\text{end}}}{M_{\text{Pl}}} \right)^{2\frac{1-\alpha}{2 - \alpha}} 
-1\right) \right), & \text{otherwise}.
\end{cases}
\end{equation}

In the hydrodynamical analysis, we found that the scale-free inflationary 
models divided into three classes, power-law-like ($\alpha \le 1$), 
intermediate ($1<\alpha < 1.5$) and plateau-like ($1.5\le \alpha$).  In the 
scalar-field potential analysis, the first class, the power-law-like 
models, divides into two cases: the strictly scale-free $\alpha=1$ case, 
corresponding to $V(\phi)= \lambda \phi^4$ with only a single {\it 
dimensionless} parameter; and $\alpha <1$, for which the potential is 
positive for all $\phi$. Both cases are free of the hydrodynamical 
initial conditions and unlikeliness problems described here and the 
corresponding problems described for potentials in \cite{Ijjas:2013vea}.   
However, in the latter case ($\alpha <1$), the potential is positive for 
all $\phi$ so there is no graceful exit unless some feature is added to 
the potential to complete it. The added feature breaks its 
appealing scale-free character. Hence, the scalar field potential 
analysis picks out the $\alpha=1$ strictly scale-free case as being 
simplest among the power-law-like class.

The intermediate class of hydrodynamical models ($1 < \alpha < 1.5$) 
translates into plateau-potentials with large-field inflation. Unlike the 
$\alpha=1$ case, these models require tuning one or more {\it 
dimensionful} parameters to satisfy cosmological constraints on the number of 
e-folds and the density fluctuation amplitude, $\delta \rho/\rho \sim 
10^{-5}$.  As in the hydrodynamical analysis, the predictions for $n_s-1$ 
and $r$ during the last 60 e-folds depend on the shape of the potential 
beyond the very flat part of the plateau as the potential dips sharply 
towards zero. Consequently, the predictions are very similar to 
expectations for monomial potentials, such as $V(\phi) \sim m^2 \phi^2$.  
However, because the potentials are plateau-like at large $\phi$, these 
models exhibit the initial conditions problem described 
here and in \cite{Ijjas:2013vea}.  

Finally, the plateau-like class of hydrodynamic models are split into two 
cases when translated into scalar-fields and potentials. For $1.5 \le 
\alpha \le 2$, they correspond to large-field models and include Higgs 
\cite{Bezrukov:2007ep} (with action expressed in Einstein frame).\footnote{We note that the estimate for $N_{\text{max}}(\alpha)$ in Figure~2 appears to be  too optimistic for $1<\alpha\le 2$ when the model is translated into quantum scalar field with potentials. For these models which have semi-infinitely long plateaus,  quantum corrections, even tiny ones, can spoil the flat plateau and block inflation.  
For example, in the case of Higgs inflation, in which the Higgs is a non-minimally coupled field and then the theory is  Weyl-transformed to the Einstein gauge,  there appears to be a semi-infinite, very flat plateau classically. This model corresponds approximately to $\alpha =2$.  However, it has been noted by several groups that this plateau at the classical level is spoiled by quantum corrections at small enough field values such that there cannot be enough inflation to solve the horizon or flatness problem \cite{Burgess:2009ea,Barbon:2009ya,Burgess:2010zq}.  In other words, with the quantum corrections, we  have $N_{\text{max}} < 60$, which is much less than the estimate in Fig.~2.   Adding extra fields can suppress the quantum corrections, but then the model is more complicated and explicitly breaks scale-freeness.}
  They 
exhibit the initial conditions and unlikeliness problems and require 
tuning one or more {dimensionful} parameters to satisfy cosmological 
constraints.  For $\alpha > 2$, the potentials correspond to      
small-field plateau-potentials such as new inflation 
\cite{Albrecht:1982wi,Linde:1981mu} which exhibit the initial conditions 
and unlikeliness problems and require two or more {\it dimensionful} 
parameters $V_{\text{end}}$ and $\phi_{\text{end}}$ to yield the correct 
spectrum of primordial density fluctuations and sufficient e-folds of 
inflation.

In sum, the model dependent analysis based on inflaton fields and 
potentials gives a somewhat different view of the landscape of scale-free 
inflationary models and their problems, but on the whole confirms and 
sharpens the results of the hydrodynamic analysis.  From either point of 
view, the strictly scale-free $\alpha=1$ case is the least problematic 
among all the models and all classes.  The analysis based on scalar 
fields with scale-free potentials splits two of the hydrodynamic classes 
into two distinct subgroups through the conversion from $N$ to $\phi$ as 
the independent variable.  It further suggests a hierarchy from least to 
most problematic, where the least problematic and requiring the least 
dimensionful parameters is the strictly scale-free $\alpha=1$ model 
followed by the intermediate class models.  Unfortunately, the 
inflationary models favored by present data does not belong to either of 
these groups.  The results also show that, in the plateau-like class, 
large-field models  with $\alpha<2$ require fewer dimensionful parameters 
than small-field models ($\alpha>2$).  

We note that the hydrodynamic unlikeliness problem  decribed in this 
paper is more general than the version identified in  
\cite{Ijjas:2013vea}. In \cite{Ijjas:2013vea} it was shown specifically 
for  small-field plateau-like models that inflation is exponentially less 
likely in a generic energy landscape than monomial potentials $V\sim 
\phi^n$.   The results in the present paper based on scale-freeness show 
that  the {\it entire} plateau-like class is theoretically disfavored 
compared to the entire power-law-like class, whether small-field or large 
field inflation.  

Among monomial inflationary potentials $V \sim \phi^n$, the only strictly 
or background-only scale-free example is the conformally invariant case, 
$n=4$, corresponding to $\alpha=1$, which we have shown is the least 
problematic.\footnote{Here we correct the crude approximation made in 
\cite{Mukhanov:2013tua} which led to the incorrect conclusion that  
$\phi^6$ is the strictly scale-free solution.}
Recovering other power-law potentials requires explicitly breaking scale-freeness while still respecting the inflationary conditions, criteria~I and II.  For example, by introducing two additional non-zero parameters 
$\beta$ and $\gamma$ as defined in Eq.~(\ref{nsf1}), the equation-of-state parameter can be made to follow closely the equation-of state that 
can be obtained  for $n=4\beta$.  Note that $\phi^2$ requires non-negligible scale-free breaking in the sense that $\beta$ is significantly less than one.  Power law models with yet smaller powers, such as $\phi^{2/3}$, require even greater deviations from scale-freeness.

However, introducing this extra scale-freeness breaking degree of freedom could be a dangerous 
course.  There already exists a spectrum of inflationary cases 
parameterized by $\alpha$ in the background scale-free limit.  Having a 
spectrum of cases reduces the predictive power of the paradigm.   
Applying the same scale-free breaking degree of freedom, $\beta$, for all 
$\alpha$ further broadens the range of possibilities and increases the 
number of parameters.  This reduces the predictability to the point where 
there can be more parameters than observational constraints.    
Furthermore, the breaking of scale-freeness only complicates the model 
without resolving any of the problems identified for the scale-free 
cases.  Given that the universe seems so simple based on observations, it 
is problematic to consider cases with more parameters than the 
inflationary paradigm requires or the data can constrain.  

Not everyone would agree with this assessment.  In order to address the 
initial conditions problem described by Ijjas et al. \cite{Ijjas:2013vea} 
and in this paper, authors have introduced potentials with double-inflation, first a power-law-like phase and then a plateau-like like 
phase \cite{Destri:2007pv,Nakayama:2013jka,Ferrara:2013rsa}; or they have 
introduced an energy landscape with false vacuum inflation tunneling to a 
plateau \cite{Vilenkin:1984wp}.   In these cases, the deviation from 
scale-freeness is intentionally designed to occur for modes outside the 
Hubble horizon beyond the range of observational tests.  From a 
theoretical perspective, the logic is odd: if the physics underlying 
inflation is not truly scale-free, why should the deviation from scale-
freeness only show up on unobservably large scales?  The only purpose is 
to evade the initial conditions problem while remaining consistent with 
observations. But the cost is too precious.  As evidenced by the example of 
Ferrara et al. \cite{Ferrara:2013rsa}, this approach introduces enough 
new parameters and enough tuning that any outcome for $n_S-1$ and $r$ 
becomes possible, such that inflationary cosmology loses all predictive 
power.   
 
\subsection{Scale-free cyclic potentials} 
 
 As explained in the Appendix,  a generic form for the scalar-field potential energy density in the cyclic model can be cast in the form:  \begin{equation}\label{V_cyclic} 
 V(\sigma, s) = V(\sigma, 0)\left(1 + \frac{1}{2}\,\kappa\,\frac{V,_{\sigma\sigma}}{V(\sigma, 0)} s^2 + \mathcal{O}(s^3)\right), 
\end{equation} 
 where $\sigma$ corresponds with the fluid component governing the background evolution described by $\epsilon_1$ and $s$ is the field representation of the fluid with equation-of-state parameter $\epsilon_2$ that generates the isocurvature fluctuations before the bounce  (that are converted to the nearly scale-invariant curvature perturbations during  the bounce).   
The background evolution is along the $\sigma$ direction with $s=0$.  The parameter $\kappa$ is the ratio $\epsilon_2/\epsilon_1$   defined in Eq.~(\ref{kappa}), which relates the curvature of the potential energy density along the $s$ direction to the curvature along the $\sigma$ direction.   The strictly scale-free case corresponds to $\kappa = 1$  such that $V,_{ss}(\sigma, s) = V,_{\sigma\sigma}(\sigma, 0)$ \cite{Buchbinder:2007tw}. 

The Friedmann equations together with Eq.~(\ref{H_cyclic}) and (\ref{e_1}) can be used to construct the potential given the background equation of state $\epsilon_1(\mathcal{N})$: \begin{eqnarray}\label{c_potential}
V(\sigma, 0) &=& -M_{\text{Pl}}^2 \left(\epsilon_1(\mathcal{N}-1) \right)H^2(\mathcal{N}) \nonumber\\ &=& -3\,M_{\text{Pl}}^2\,H^2_{\text{end}} \left((\mathcal{N}+1)^{\alpha_1}-1 \right) \exp\left( - 2\,\mathcal{N} + 2\int^0_{\mathcal{N}}\frac{d\mathcal{N}}{3(\mathcal{N}+1)^{\alpha}-1}
\right)
,
\end{eqnarray}
 where $\mathcal{N}$ can be replaced by the background scalar field $\sigma$ using the relation  \begin{eqnarray}\label{sigma} \frac{\sigma - \sigma_{\text{end}}}{M_{\text{Pl}}} & = & \pm \int^0_{\mathcal{N}} \sqrt{2\epsilon_1}\,(\epsilon_1-1)^{-1}d \mathcal{N} \nonumber\\ & = &  \pm \sqrt{6} \int^0_{\mathcal{N}} \frac{(\mathcal{N}+1)^{\alpha_1/2} }{3(\mathcal{N}+1)^{\alpha_1}-1}d \mathcal{N} .
\end{eqnarray}
For example, for $\alpha_1 = 1$ we have
\begin{equation}\label{example}
V(\sigma,0) \simeq -3\, M_{\text{Pl}}^2\,H^2_{\text{end}} \left(  \sigma^2/M_{\text{Pl}}^2 - 1 \right)\exp\left( -2 \sigma^2/M_{\text{Pl}}^2 \right).
\end{equation}
Here we set without loss of generality $\sigma_{\text{end}} = 1$ and assumed $\sigma - \sigma_{\text{end}} >0$ during the smoothing phase.  
For all $\alpha>0$, the potential $V(\sigma, 0)$ takes the same generic form:  a steep negative potential  that reaches a minimum  before $\sigma$ approaches $\sigma_{\text{end}}$, the standard shape potential proposed for ekpyrotic and cyclic scenarios.  (This can be checked by computing the derivative of Eq.~(\ref{c_potential}), $d\,V/d\,\mathcal{N}$ for different $\alpha$ and by observing
from Eq.~(\ref{sigma}) that the transformation from ${\mathcal N}$ to $\sigma$, $\mathcal{N}(\sigma)$, is strictly monotonic.)   

This means that  the potential picture gives the same simple result as the model independent hydrodynamic analysis, namely that  the scale-free cyclic theory  has only a single class of models all requiring a single dimensionful parameter, $H^2_{\text{end}}$, to yield the correct spectrum of primordial density fluctuations, $\delta\rho/\rho \sim 10^{-5}$.  Hence, both pictures lead to the conclusion that there is no unlikeliness problem and no extra parameters or fine-tuning problem can arise. 
 
\section{Discussion}

In this paper, our aim has been to study different cosmological scenarios 
in a model independent way that does not refer directly to fields or 
potentials. Using a hydrodynamic approach, we derived algebraic forms for 
the equation-of-state parameter consistent with the scale-free principle 
for both inflationary and cyclic theory. In this section  we discuss both 
theoretical and observational implications of this work.

Let us first consider inflationary cosmology alone.  We found that, based 
on our hydrodynamical analysis, inflationary scale-free models divide 
into three distinct classes and identified a range of related problems: 
an initial conditions problem for the plateau-like and intermediate 
class, and an unlikeliness problem and a fine-tuning problem for the 
plateau-like class. The spectrum becomes even more divided when we 
translate the three cases into scalar-field potentials.  Hence, even 
limiting ourselves to scale-freeness, there is a diversity of 
inflationary models and predictions.

In applying the same hydrodynamic analysis to cyclic scenarios, we found 
cyclic theory allows only a single scale-free class of models and does 
not suffer from the initial conditions or unlikeliness-type problems 
identified for inflation.  
At the perturbative level, current versions of 
cyclic theory require a two-component fluid for the generation of 
primordial isocurvature fluctuations, which are then converted into 
density fluctuations. This added condition compared to inflation appears 
to have no disadvantage in a hydrodynamical treatment assuming scale-freeness:  there were no more parameters, fine-tuning, or other kinds of 
constraints compared to the inflationary one-fluid mechanism. Remarkably, translating this single cyclic class into scalar-field potentials, we found the same simple result.

One might ask if the problems found for inflation that were not found for 
cyclic may be related to the fact that a single fluid was assumed in the 
first case but not the second.  The answer is no.  
As we discussed above in \S~II, in scale-free scenarios the background is 
always described by a single fluid component and the presence of multiple 
components becomes relevant only at perturbative level. However, the 
inflationary problems arise at background level such that adding multiple 
fluid components makes (at best) no difference whatsoever. In fact, the   
situation for inflation is typically made worse.
For example, there is a well-known two-component fluid version of 
inflationary theory, known as the curvaton model \cite{Lyth:2001nq}. As 
in the cyclic model, the background evolution is governed by one fluid 
component, the inflaton, and the perturbations are controlled by another, 
the curvaton.   Since the inflaton must satisfy the same conditions on 
the equation-of-state as in the single-fluid case, there is no change 
whatsoever in the problems encountered by introducing the curvaton.  
Since both fluids are capable of generating density perturbations,  extra 
fine-tuning is required to regulate the interplay of the inflaton and 
curvaton in order that  only  the curvaton affects the evolution of 
perturbations.   That is, a curvaton is not automatically the leading 
order contributor to the perturbations; the model must be adjusted to 
make it so.  In particular the curvaton construction requires setting 
$\epsilon_1(N)$ for the inflaton different from $\epsilon_2(N)$ for the 
curvaton, which explicitly breaks background scale-freeness.  This is 
qualitatively different from the cyclic case where two fluids are 
required to generate the leading order contribution to the density 
perturbations and $\epsilon_1(\mathcal{N})$ can be set equal to 
$\epsilon_2(\mathcal{N})$, preserving  scale-freeness, as was done in \S 
IV.B.

Finally, we relate our theoretical findings to current observations, in 
particular to recent Planck satellite measurements \cite{Ade:2013rta}.  
We see that strictly scale-free versions of both cosmological scenarios 
are observationally disfavored. The strictly scale-free $\phi^4$-chaotic 
inflation potential is observationally disfavored by more than $4\sigma$ 
as a result of constraints on $n_S$ and $r$. The strictly scale-free 
cyclic model is consistent with current bounds on $r$ but predicts $n_S-1 
\simeq - .01$, which is disfavored by $3\sigma$.  
That means, consistency with current observational data requires some 
deviation from strict scale-freeness  in both scenarios. 

In the cyclic theory the observational value of $n_S-1$ can be obtained 
simply by introducing a very weak breaking of scale-freeness at the 
perturbative level ($\beta_2$ slightly different from 1 or, equivalently, 
$|\kappa-1|\ll1$),  while leaving the dominant fluid and the background 
strictly scale-free ($\beta_1=1$).  
In inflation, by contrast, the current observations favor scale-freeness 
only for plateau-like models, which suffer from the initial conditions 
and unlikeliness problems described above. The only power-law-like models 
that are not strongly disfavored require significant breaking of scale-freeness ($|\beta-1| \sim \mathcal{O}(1)$).

\begin{table*}[t]
\begin{center}
\renewcommand\tabcolsep{8pt}
\renewcommand{\arraystretch}{1.2}
\begin{tabular}{|c|c|c|c|}\hline
\multirow{2}{*}{$r$} & \multirow{2}{*}{$n_S-1$}  
&\multirow{2}{*}{\parbox{.75in}{ unlikeliness problem}}& 
\multirow{2}{*}{favored model}\\
 &&&\\\hline
 \multirow{3}{*}{\parbox{.6in}{$\gtrsim10^{-4}$}} &  
\multirow{2}{*}{\parbox{1.7in}{scale-free satisfying Eq.~(\ref{nsf2}) 
with $|\beta -1| \ll 1$}} 
& no, if $r\gtrsim 0.1$& \multirow{2}{*}{\parbox{.75in}{scale-free 
inflation}}\\\cline{3-3}
 &     &  yes, if $0.1\gtrsim r\gtrsim 10^{-4}$ ${}^{\ast}$ &\\\cline{2-
4}
  &violating Eq.~(\ref{nsf2})   & \multicolumn{2}{c|}{?} \\\hline
\multirow{2}{*}{\parbox{.6in}{$\lesssim 10^{-4}$}}& 
\multirow{2}{*}{\parbox{1.7in}{scale-free satisfying 
Eq.~(\ref{n_s_cyclic}) with $|\kappa -1| \ll 1$}} &  no & scale-free 
cyclic theory\\\cline{3-4}
 &  &   \multicolumn{2}{c|}{?} \\\hline
\end{tabular}
\end{center}
\label{default}
\caption{Testing scale-free primordial cosmology with measurements of the 
tensor-to-scalar ratio $r$ and the tilt $n_s-1$. See discussion in text.  
${}^{\ast}$Note that the results from our model-dependent analysis in 
\S~V based on scalar fields and potentials further divide plateau-like 
models into two groups:  $\alpha \leqslant 2$, which requires $r \gtrsim 
0.004$; and,  $2 < \alpha$, which requires $10^{-4} \lesssim r \lesssim 
0.004$, where this latter group requires more dimensionful parameters and 
has a more severe unlikeliness problem.}
\end{table*}

What will future observations tell us about scale-free primordial 
cosmology? Scale-free inflation is already in serious jeopardy given what 
we know:  there are the historic entropy 
\cite{Penrose:1988mg,Gibbons:2006pa} and multiverse 
\cite{Steinhardt:1982kg,Vilenkin:1983xq} problems that apply to {\it all} 
inflationary models \cite{Steinhardt:2011zza}. Hence, at best, we have 
these problems to overcome. However, future observations could make 
matters worse for scale-free inflation. 
We summarize all possible scenarios in Table~1. 

An important {\it prediction} for scale-free inflation that stems from 
this work is that the tensor-to-scalar ratio $r$ should exceed 0.0001, 
which is within conceivable experimental sensitivity. (Here, as throughout the paper, we assume $c_s >1/3$, as implied by current observations \cite{Ade:2013ydc}.)    This bound arises 
because smaller $r$ requires $\alpha > 3$, which, in turn, requires $n_S 
<0.95$ in disagreement with current measurements of the spectral tilt.  Note that  the tensor-to-scalar ratio, $r$, does not depend on the energy scale of inflation since it precisely cancels from the ratio. Models with $r$ far below $10^{-4}$ either violate existing observational constraints (such as the limit on $n_S-1$) and/or introduce extra parameters that strongly break scale-freeness.
If none of the scale-free combinations of ($r, n_S-1$) is found observationally, scale-free inflation is ruled out.   If one of these combinations is observed with $10^{-4}< r \lesssim 0.1$, then scale-free inflation is possible, but it is necessary to resolve the initial conditions and unlikeliness problems discussed here.   If a combination is found with $r>0.1$, scale-free inflation without either of these problems is possible (though there would remain the entropy and multiverse problems common to all inflationary models).   

The current situation is that observations indicate $r < 0.1$.  Hence, unless future B-mode measurements bring a surprise that overrules this result, the only possible scale-free inflationary models remaining encounter the initial conditions and unlikeliness problems discussed here. 

Alternatively, future observations could find that the measured values of 
$r$ and $n_S-1$ yield no scale-free combination consistent with 
Eq.~(\ref{nsf2}), or
$r<0.0001$.   
Either case would eliminate all scale-free inflationary models and force 
extra degrees of freedom that allow virtually any outcome for $n_S-1$ and 
$r$, as exemplified by the scale-freeness violating model of Ferrara et 
al. \cite{Ferrara:2013rsa}. In this case, inflationary cosmology loses 
all predictive power.    

As for scale-free cyclic models, the situation is somewhat different.  
There is no multiverse problem and the initial conditions and 
unlikeliness problems found for inflation are evaded.  Observationally, 
the strictly scale-free cyclic case ($\alpha=1$) is disfavored because of 
the current constraints on the spectral tilt. 
A best fit to the tilt requires a small deviation from scale-freeness at 
the perturbative level, by setting  $\beta_2$ (or, equivalently $\kappa$) 
slightly greater than 1 instead of equal to 1 precisely. 
The forthcoming measurements of $r$ are crucial to scale-free cyclic 
models because all predict no observable tensor modes. Detection of 
primordial gravitational waves would eliminate the entire spectrum of 
models. On the other hand, if there is no detection and $r$ is proven to 
be less than 0.0001 -- the  conditions that eliminates scale-free 
inflation -- scale-free cyclic would fit perfectly. 

In the cyclic models considered here, we have assumed an entropic 
mechanism with two fluids for generating curvature perturbations.  At 
least in currently known examples in which this is achieved with two 
scalar fields, the models generate non-negligible $f_{\text{NL}}$ or 
$g_{\text{NL}}$ or both.  Current observational limits are consistent 
with predictions without requiring any additional tuning of parameters 
\cite{Lehners:2013cka}, but future measurements could result in detection 
or tighter constraints.  Although non-Gaussianity is not directly 
predicted by hydrodynamical analysis and is more model-dependent in 
cyclic models, future measurements could be useful in distinguishing 
inflation versus cyclic scenarios and the testing the hypothesis of 
scale-free primordial cosmology.

In sum, introducing the scale-free principle makes cosmological theories 
-- both inflationary and cyclic -- meaningfully predictive and allows for 
observational test. Both for scale-free inflationary and cyclic 
cosmology, we could identify all combinations of parameters $(r, n_S-1)$ 
consistent with the theory.  If such a combination  is not measured, the 
theory is falsified. Most interestingly, forthcoming measurements are 
capable of testing and eliminating scale-free inflationary models, scale-
free cyclic models, or both, as indicated by the ``?'' in Table~I.  
Eliminating both means relinquishing scale-freeness and  having to settle 
for unpredictive theory, like \cite{Ferrara:2013rsa}, or seeking another 
type of cosmological theory that retains scale-freeness and predictive 
power.

\begin{acknowledgments}
This work was inspired in part by a talk and recent paper by V. Mukhanov 
\cite{ Mukhanov:2013tua}.   We thank J. Khoury, J.-L. Lehners, and N. 
Turok for very helpful comments on the manuscript.
This research was partially supported by the U.S. Department of Energy 
under grant number DE-FG02- 91ER40671 (PJS), by NSF grant AST-0907890 and NASA grants NNX08AL43G and NNA09DBB30A (AL), and
the Perimeter Institute for Theoretical Physics (AI and PJS). Research at 
Perimeter Institute is supported by the Government of Canada through 
Industry Canada and by the Province of Ontario through the Ministry of 
Research and Innovation.  AI also acknowledges the support of the 
European Research Council via the Starting Grant No.~256994. The work of 
AI is supported in part by a grant from the John Templeton Foundation. 
The opinions expressed in this publication are those of the authors and 
do not necessarily reflect the views of the John Templeton Foundation. 
AI thanks the Physics Department of Princeton University for hospitality while this research was completed.

\end{acknowledgments}

\appendix*
\section{Derivation of Eq.~(\ref{n_s_cyclic0})}

In order to derive the general hydrodynamic expression for the spectral tilt of primordial density fluctuations in cyclic theories, we follow the same procedure as for inflation \cite{Wang:1997cw}. Namely, we first solve for the perturbations, assuming the fluids can be represented as scalar fields with potentials, and then we convert the potential parameters in the expression derived for the tilt into hydrodynamic variables.
To represent the two-component fluid we choose two fields, $\sigma$ and $s$, where $\sigma$ corresponds to the fluid component governing the background evolution described by equation of state $\epsilon_1$ and $s$ is the field representing  the fluid that generates the isocurvature fluctuations  before the bounce that are later converted to curvature perturbations during the bounce.  The second fluid has equation-of-state parameter $\epsilon_2$. 
The perturbation equation is given by
\begin{equation}\label{pert_c}
\delta\ddot{s} + 3H\delta\dot{s} + \left( \frac{k^2}{a^2} + V,_{ss} \right)\delta s= 0\,, \end{equation} where dot denotes derivation with respect to physical time, $k$ is the adiabatic mode. 

For the cyclic potential we choose the form \begin{equation}\label{V_cyclic} V(\sigma, s) = V(\sigma, 0)\left(1 + \frac{1}{2}\,\kappa\,\frac{V,_{\sigma\sigma}}{V(\sigma, 0)} s^2 + \mathcal{O}(s^3)\right), \end{equation} in agreement with \cite{Buchbinder:2007tw,Lehners:2013cka}.
Here $\kappa$ is
the ratio of the equation-of-state parameters , $\kappa \equiv  \epsilon_2/ \epsilon_1$ as  in Eq.~(\ref{kappa}).   $V(\sigma, s)$ is constructed such that for $\kappa=1$ it yields scale-free solutions; this corresponds to the case $V(\sigma,s),_{ss}=V(\sigma,0),_{\sigma\sigma}$. 
Parameterizing the cyclic potential in this way is useful since the form naturally incorporates the entropic mechanism by  dividing  the potential into a first factor, that describes the background evolution along the $\sigma$ direction and the second factor, which describes the direction of  the isocurvature perturbations.   Furthermore, this form encompasses all known simple cyclic potentials, such as models that can be written as sums of exponentials of independent fields. 

After rescaling $\delta S \equiv a(\eta) \delta s$ and assuming standard 
Bunch-Davies initial conditions, $\delta s \rightarrow e^{-
ik\eta}/(2k)^{3/2}$, the solution of Eq.~(\ref{pert_c}) is the Hankel 
function 
\begin{equation}
\delta s = \frac{\sqrt{-\pi\eta}}{2}H^{(1)}_{\nu}(-k\eta)\,,
\end{equation}
with
\begin{equation}\label{nu}
\nu^2 = \frac{1}{4} + \eta^2\left( \frac{a''}{a} - a^2\kappa\, 
V,_{\sigma\sigma} \right)
\,.
\end{equation}
Here prime denotes derivative with respect to conformal time $\eta$.
On large scales, $k\ll aH$, $\delta s \sim k^{-\nu}$. This corresponds to 
a spectral tilt
\begin{equation}\label{tiltexp}
n_S-1= 3 - 2\nu.
\end{equation}
To express the tilt in hydrodynamical language, we follow 
\cite{Lehners:2007ac} and rewrite $H, a,$ and $V,_{\sigma \sigma}$ in 
terms of the background equation-of-state parameter 
$\epsilon_1(\mathcal{N})$,
\begin{eqnarray}
(a\,H)^{-1} & \simeq & \epsilon_1 \eta \left( 1 - \frac{1}{\epsilon_1} - 
\frac{\epsilon_1,_{\mathcal{N}}}{\epsilon_1} \right) ,\\
\frac{a''}{a}& \simeq & 2\,a^2H^2 \left( 1 - \frac{1}{2\,\epsilon_1} 
\right),\\
V,_{\sigma \sigma} &\simeq& -H^2 \left( 2\epsilon_1^2 - 6\epsilon_1 - 
\frac{5}{2}(\epsilon_1 -1)\epsilon_1,_{\mathcal{N}}\right) .
\end{eqnarray}
After some algebra, we find 
\begin{equation}
\nu^2 \simeq \frac{1}{4} + 2 \left( \kappa + 3\cdot \frac{1-
2\kappa}{2\epsilon} + \frac{8-5\kappa}{4} \cdot 
\frac{\epsilon,_{\mathcal{N}}}{\epsilon} \right)\,,
\end{equation}
where we neglected terms of order $1/\epsilon^2$.
Finally, substituting into Eq.~(\ref{tiltexp}) yields the hydrodynamic 
expression for the spectral tilt as stated in Eq.~(\ref{n_s_cyclic0}).

\bibliographystyle{apsrev}

\bibliography{slava_refs}

\end{document}